\documentclass[a4paper]{article}

\usepackage{atmohead2013}
\usepackage[english]{babel}

\title{ARCADE - Atmospheric Research for Climate and Astroparticle DEtection}

\shorttitle{The ARCADE Project}

\authors{
M. Buscemi$^{1}$,
C. Cassardo$^{2}$,
M. Cilmo$^{1}$,
M. Coco$^{2}$,
S. Ferrarese$^{2}$,
F. Guarino$^{1}$,
A. S. Tonachini$^{2}$,
L. Valore$^{1}$,
L. Wiencke$^{3}$
}

\afiliations{
$^1$ University of Naples ”Federico II” and INFN - Naples, Italy\\
$^2$ University of Turin and INFN - Turin, Italy\\
$^3$ Colorado School of Mines, Golden, Colorado - USA\\

}

\email{mario.buscemi@na.infn.it}

\abstract{The characterization of the optical properties of the atmosphere in the near UV, in particular the tropospheric aerosol stratification, clouds optical depth and spatial distribution are common in the field of atmospheric  physics, due to aerosol effect on climate, and also in cosmic rays physics, for a correct reconstruction of energy and longitudinal development of showers. The goal of the ARCADE project is the comparison of the aerosol attenuation measurements obtained with the typical techniques used in cosmic ray experiments (side-scattering measurement, elastic LIDAR and Raman LIDAR) in order to assess the systematic errors affecting each method providing simultaneous observations of the same air mass with different techniques. For this purpose we projected a LIDAR that is now under construction: it will use a 355 nm Nd:YAG laser and will collect the elastic and the N$_2$ Raman back-scattered light. For the side-scattering measurement we will use the Atmospheric Monitoring Telescope, a facility owned by the Colorado School of Mines and placed in Lamar (Colorado), the site where our experiment will take place.}


\begin{document}
\maketitle

\section{Introduction}

The knowledge of the characteristics of atmospheric aerosols is an important topic in different scientific fields. It is closely related to the understanding of the complex mechanisms that determine the climatic variations of our planet, and nevertheless aerosols and clouds play a key role in the observations and study of High Energy Cosmic Rays ($E > 10^{12} eV$) and Ultra High Energy Cosmic Rays (UHECR) ($E > 10^{18} eV$) from both ground-based and space-based experiments \cite{bib:atmo, bib:space}. \\
Common techniques adopted to observe extensive air showers (EAS) are based on the detection of the Cherenkov light \cite{bib:CTA} and UV fluorescence light \cite{bib:auger} that is emitted during the passage of secondary particles of the shower through the atmosphere. These techniques, that are sensitive to the development of cosmic ray showers in atmosphere, allow the identification of the primary particle that generated the shower and the reconstruction of its energy and mass composition from the study of the development of the light profile.\\
The fluorescence technique to detect EAS makes use of the atmosphere as a giant calorimeter whose properties must be continuously monitored to ensure a reliable energy estimate. Atmospheric parameters influence both the production of fluorescence light and its attenuation towards the detectors. The molecular and aerosol scattering processes that contribute to the overall attenuation of light in the atmosphere can be treated separately.  Molecular scattering is well described by the Rayleigh theory, and can be estimated by measuring atmospheric state variables. The scenario is more complex for aerosol attenuation, it depends on the particulate composition, size, and shape, and in general it can't be calculated analytically.  Aerosols are subject to significant variations on time scales of an hour and aerosol attenuation is the largest time dependent correction applied for air shower reconstruction. Due to the importance of this aspect of the atmospheric properties, EAS observatories need to develop ad-hoc atmospheric monitoring systems.

The ARCADE project aims to characterize the optical properties of the atmosphere in the near UV region and to develop predictive models through measurement campaigns in areas with different environmental characteristics. In detail, for the first time, different techniques of analysis used in Cosmic Rays experiments to study aerosol attenuation, cloud cover and cloud optical depth will be applied simultaneously on the same air mass to highlight the points of strength, limitations and specially reduce the systematic errors that affect each technique. We also plan to develop models that describe the stratification of aerosols in the areas of measurement and their dependence on temperature, wind, precipitation, humidity, in collaboration with climatology experts that will apply predictive models to these measurements.

\section{Measurement of aerosol attenuation using laser-light scattering}\label{method}
The use of laser sources is fundamental to obtain fast vertically resolved measurements of the optical property of the atmosphere such as aerosol attenuation. Presently the typical techniques used  in cosmic ray observatories to study aerosol attenuation are based on the use of elastic LIDARs and on the analysis of the side-scattered laser light collected by a far UV detector. Both techniques make use of a laser beam fired through the atmosphere and consist in the detection of scattered light through a telescope. In ARCADE we will make use of a steerable Raman LIDAR and of a UV detector for side scattering measurements (see sect.~\ref{apparatus}). 

Analytically the amount of light collected by the detector depends on the light intensity at the source $I_0(\lambda,s)$ and is a function of the molecular $T_{mol}(\lambda,s)$ and aerosol $T_{aer}(\lambda,s)$ transmission coefficient of the atmosphere along the track of the light:
\begin{equation}
I(\lambda,s)=I_0(\lambda,s)T_{mol}(\lambda,s)T_{aer}(\lambda,s)(1+H.O.)\frac{d\Omega}{4\pi}
\end{equation}
$T_{mol}(\lambda,s)$ will be evaluated from GDAS model \cite{bib:gdas} and, assuming a horizontally uniform aerosol distribution, $T_{aer}(\lambda,s)$ is a simple function of the aerosol extinction coefficient, $\alpha_{aer}(\lambda,s)$, or of the vertical aerosol optical depth~(VAOD):

\begin{equation}
T_{aer}(\lambda,s)=e^{-\int{\alpha_{aer}(\lambda,s)ds}}=e^{-\frac{VAOD(h)}{sen(\phi)}}
\end{equation}
Where $\phi$ is the elevation angl oh the light path.\\
Side scattering data will be analyzed with the two different procedures currently adopted by the Pierre Auger collaboration \cite{bib:clf}. 
The Data Normalized method determines the aerosol attenuation coefficient $\alpha_{aer}$ by comparing measured profiles to a reference profile chosen in a night in which the aerosol attenuation is negligible.\\
The Laser Simulation method estimates $\alpha_{aer}$ by the comparison of measured profile with a grid of profiles simulated varying the aerosol  attenuation, described by a two parameters model, using different values of the aerosol horizontal attenuation length $L_{aer}$ and of the aerosol scale height $H_{aer}$. The first describes the light attenuation due to aerosols at ground level, the latter accounts for its dependence on height. The Laser Simulation method will be also refined by introducing in the simulation a third parameter, the mixing layer height $H_{mix}$, that takes into account the planetary boundary layer (PBL). The vertical aerosol optical depth can be expressed as a function of these three parameters as follows:

\begin{equation}
VAOD(h)=\frac{H_{mix}}{L_{aer}}+ \frac{H_{aer}}{L_{aer}} \left(   1 - e^{-\frac{(h -H_{mix})}{H_{aer}}}  \right)
\end{equation}

Another instrument often adopted in many observatory to measure aerosol attenuation is the elastic LIDAR \cite{bib:LIDAR}. The amount of light collected by the LIDAR is related to the laser power $P_0(\lambda)$ by the so-called LIDAR equation:
\begin{equation}
P(\lambda,z)=P_0(\lambda)\frac{K(\lambda)}{z^2} \beta(\lambda,z)   e^{-\int{\alpha(\lambda,z)dz}}
\end{equation}
where the coefficient $K(\lambda)$ takes into account all geometrical factors, $\beta(\lambda,z)$ is the back-scattering coefficient, and $ \alpha(\lambda,z)$ is the sum of the aerosol and molecular attenuation coefficients.
One of the simplest way to solve this equation is the Klett method \cite{bib:klett}, but this technique is based on the assumpion of a power law relationship between backscatter and attenuation coefficients, assumption that makes this method weak.\\ 
A better solution can be achieved by exploiting the use of a steerable LIDAR: the multiangle analysis estimate the aerosol attenuation from the ratio of the signal collected by the LIDAR at various tilted angles.
The weakness of this technique is that it depends on the assumption of a perfect horizontal uniformity of the optical properties of the atmosphere, a property that is not always verified.

Another method that overcomes these limitations consists in the measurement with a LIDAR of both elastic and Raman-shifted back-scattered light: the greatest drawback of this technique is that the Raman scattering cross section is much smaller ($\sim$ 10$^{-3}$) than elastic one, so a consistent measurement needs a powerful laser source and very long time for acquisition. Due to this feature Raman LIDARs have never been used in the field of view of cosmic ray telescopes during air shower acquisition until now, but only at the beginning and at the end of the shift. Moreover, if only one shifted wavelenght is measured,  this method need to assume an analytical dependence of the aerosol attenuation coefficient at different wavelengths $\lambda_0$ and $\lambda_R$.

Comparing all these techniques ARCADE will clarify the limit of applicability that these assumptions introduce in each method.

\section{Measurement Apparatus} \label{apparatus}
\begin{figure}[h]
  \centering
\includegraphics[ width=0.45\textwidth]{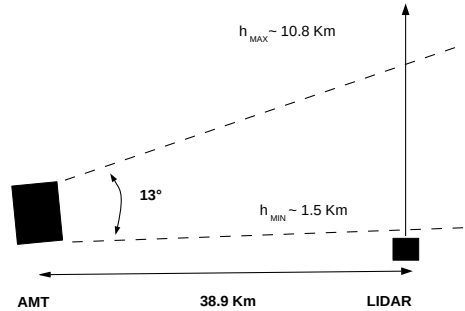}
  \caption{ Illustration of the side view of the Raman LIDAR and side scattering aerosol measurement system. The
distance of the AMT and Raman LIDAR is 38.9 km. The
AMT is tilted by 2.2$^{\circ}$ in elevation, and has a field of view
of 13$^{\circ}$.}
  \label{setup}
 \end{figure}

The experiment will take place in a desertic zone in southeastern Colorado. The site is a flat plateau at about 1200 m a.s.l. with low air pollution: these characteristics make the site suitable for cosmic ray observations. The measurement apparatus consists in a steerable Raman LIDAR and in a UV telescope placed at about 40 km from LIDAR, that will collect the light emitted by the same laser source of the LIDAR and side-scattered by the atmosphere (Fig.~\ref{setup}).

\subsection{The Atmospheric Monitoring Telescope}
\begin{figure}[t]
  \centering
  \includegraphics[width=0.3\textwidth]{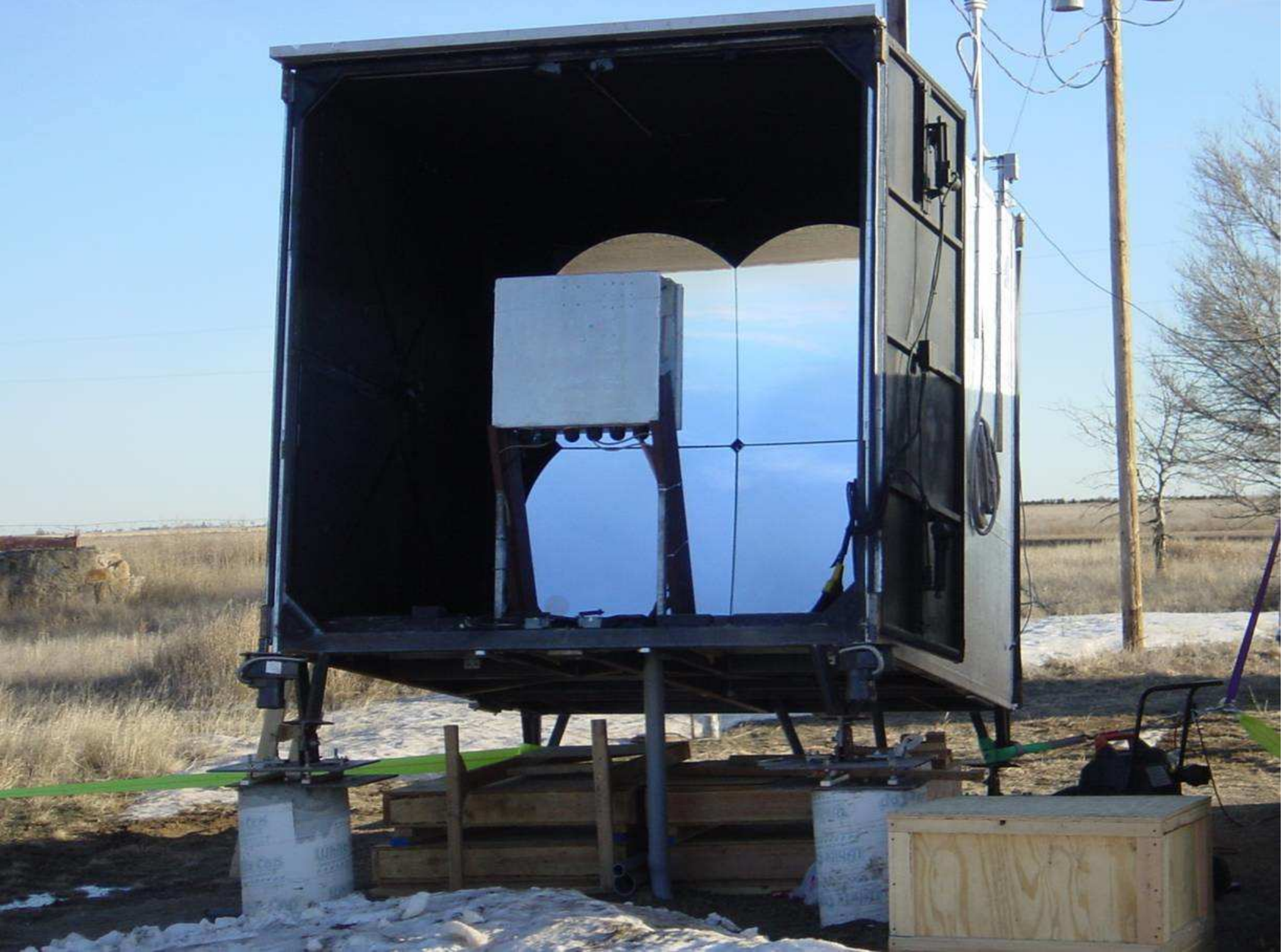}
  \includegraphics[width=0.3\textwidth]{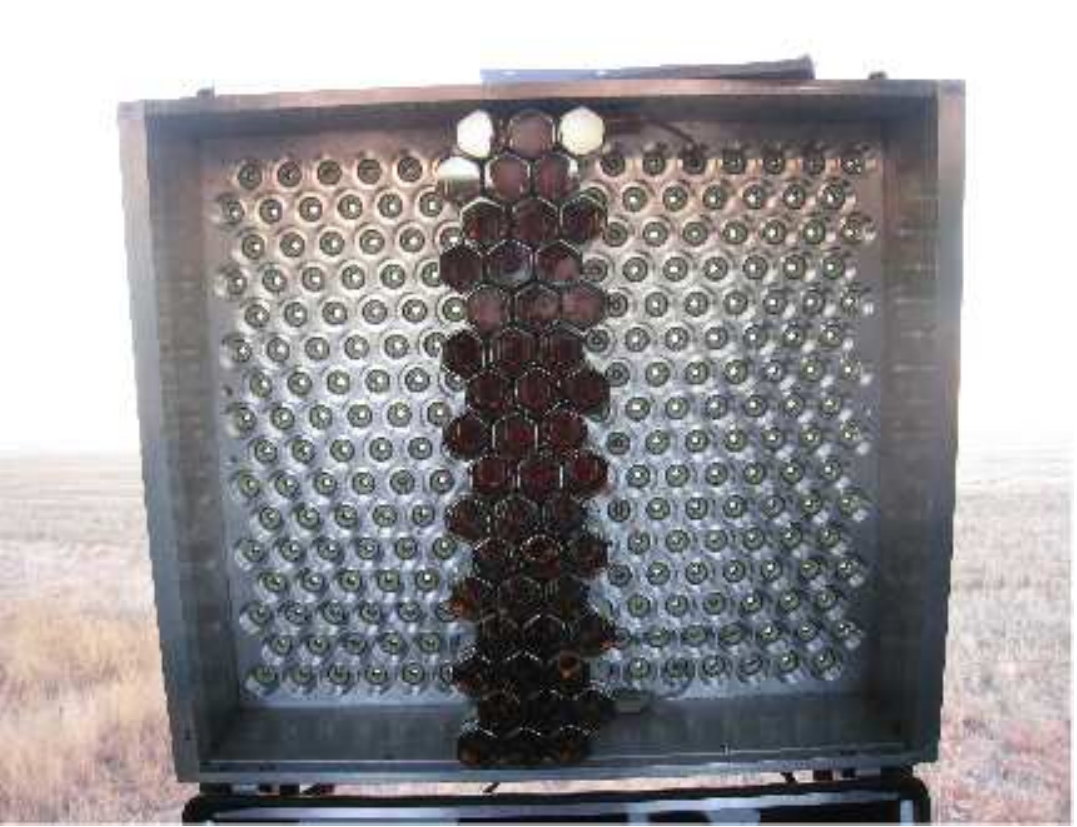}
  \caption{The AMT at the Two Buttes site, Colorado. Top :
the container, the 4-segment spherical mirror and the back
of the camera are visible. Bottom : front of the AMT camera,
with the 3 columns of 16 PMTs.
}
  \label{camera}
 \end{figure}

The Atmospheric Monitoring Telescope (AMT) is the facility that will be used for the side-scattering measurement: it is a telescope for the detection of UV light and it is owned by the Colorado School of Mines. The telescope was built during 2008 with spares recovered from the HiRes experiment for the R$\&$D of the Auger North experiment, and is composed of a 4-segment spherical mirror having a total area of 3.5 m$^2$, and of a camera, placed in its focal plane, equipped with three columns of sixteen Photonis XP3062 photomultipliers with hexagonal window and a field of view of 1$^{\circ}$ (Fig.~\ref{camera}). A UV filter centered at 355 nm is placed in front of the camera to reduce background light. 

The AMT is placed in the region near Two Buttes at about 40 km from the laser source of the LIDAR and will measure the laser light, fired at low repetion rate (4 Hz), scattered at large angles towards the telescope. The PMT readout is based on data acquisition electronics designed for the HEAT telescopes of the Pierre Auger Observatory \cite{bib:heat}.
The sampling rate of the digitizing system is 20 MHz. The readout is triggered externally, either by a GPS device in
order to synchronize the acquisition with laser firing, or from the UV LED system used for calibration. The AMT
is housed in a dedicated waterproof container, with an automated roll-up door in front. Pillars and adjustable feet
were constructed to obtain an angle of elevation of AMT of 8.72$^\circ$ from the horizontal.

The camera of the AMT is currently at the Colorado School of Mines for testing and improvements. The data acquisition system has been recently upgraded, and the new calibration system and software provided by Colorado State University is now ready. Once fully tested in the laboratory, the camera will be transported to the Two Buttes site for installation, alignment and first light collection. A weather station will be located close to the container of the AMT to monitor wind speed, rain, temperature and humidity of the site in real time. A single board computer housed in the back of the AMT container will record these informations, and in case of poor weather conditions the computer will automatically close the doors to protect the equipment inside the AMT.

\subsection{The Raman LIDAR}
\begin{figure}[t]
  \centering
  \includegraphics[width=0.3\textwidth]{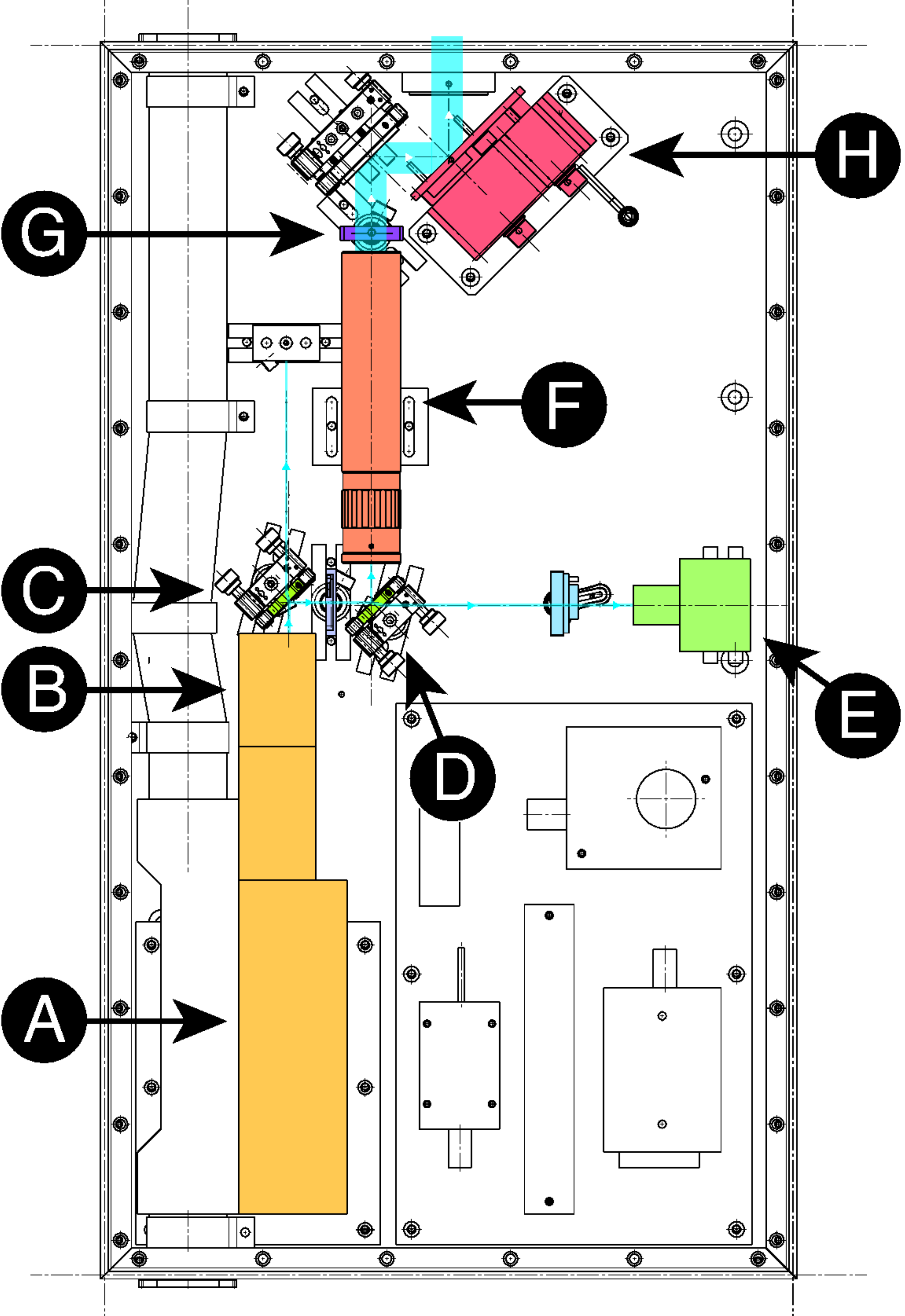}
  \caption{Laser bench scheme. The laser (A) emits $\o\,$1.6\,mm light beam with a full divergence of $\sim\,3$\,mrad. Light is purified by five dichroic mirrors (B,C). A beam splitter (D) sends 5\% of the light to a laser probe (E). The main beam passes though a 10X beam expander (F), thus reducing the divergence to $\sim\,0.3$\,mrad. The light is then depolarized (G). The beam alignment can be finely controlled with a motorized mirror mount (H). The beam finally exits the laser box passing through a quartz window.}
  \label{bench}
 \end{figure}

\begin{figure}[t]
  \centering
  \includegraphics[width=0.3\textwidth]{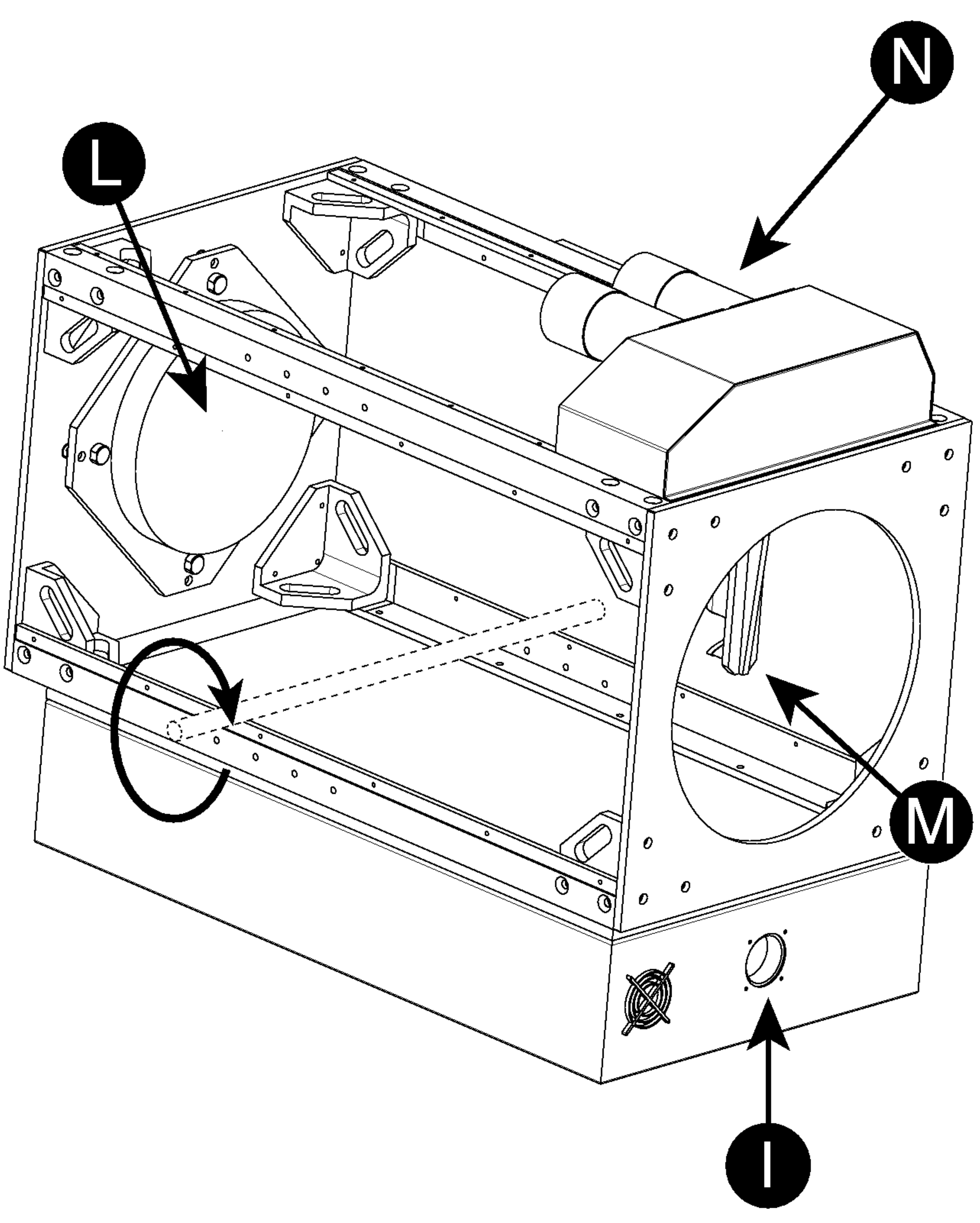}
  \caption{LIDAR scheme. The laser beam exists from (I). The backscattered light is collected by a primary parabolic mirror (L) and sent with a secondary flat mirror and a lens (M) into the Raman box (N), where the different wavelengths are split and sent different PMTs. The steering axis is represented by a dashed cylinder. }
  \label{LIDAR}
 \end{figure}

A new steerable Raman LIDAR was designed for the purpose of the project, and now it is in construction. The laser source of the device is a Quantel Centurion Nd:YAG diode-pumped solid state laser with a second and third harmonic generation module that emits UV light at $\lambda_0$~=~355 nm with residual at 1064 and 532 nm. Light at these wavelenghts is filtered trough a system of five dichroic mirrors that guarantee an ultra pure beam at $\sim$~6 mJ. A sketch of the optical bench is shown in Fig.~\ref{bench}. An RjP-445 pyroelectric energy probe monitors the laser power measuring 5\% of the beam output: this measurement is needed for the side-scattering analysis. A 10X beam expander reduces the laser beam divergence down to 0.3 mrad, and a depolarizer makes the light polarization almost random to avoid a preferencial direction in the scattering process.
Before the laser beam exits the box, a two-axis motorized mirror mount, holding a 2$''$ flat mirror and controlled by computer, allows fine adjustments of its output direction. A $\phi$ 25 cm primary parabolic mirror collects the backscattered light and reflects it on a flat mirror that deflects the light into the Raman box (Fig.~\ref{LIDAR}) . The return light is thus launched on a beam splitter which separates the elastic and Raman backscattered photons. After the beam splitter a 354.7 $\pm$ 2 nm narrowband filter received the resulting elastic beam, while the Raman-shifted line from nitrogen ($\lambda_R$ = 387 nm) is selected by another narrowband filter. 


Each of the two beams is read by a different 2$''$ R1332 photomultiplier tubes by Hamamatsu  and output signals are amplified and sampled with a 10 bit 2 GS/s digitizer.

The parallax between the laser beam and the receiver is 301 mm. 
From ray tracing simulations it is expected to be able to have a complete overlap of the laser beam and 
the mirror field of view at a distance of about 250 m.
The LIDAR is steerable, and can be pointed in any direction between 
zenith and the horizontal towards the AMT. The movement is handled by 
a Trio MC224 motion controller, which operates a servo motor, and reads the actual 
position of the telescope from a 13 bits absolute encoder directly mounted on the steering axis.

\section{Operation and analysis}

\begin{figure}[t]
  \centering
  \includegraphics[width=0.45\textwidth]{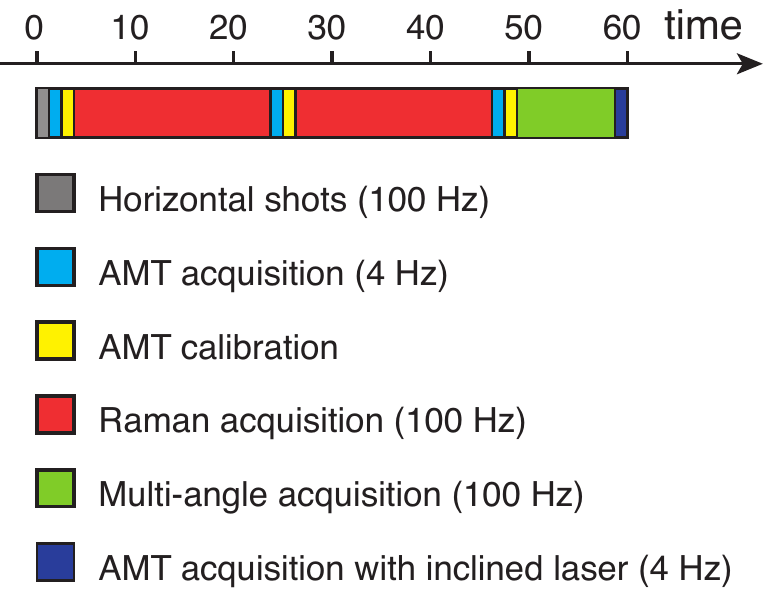}
  \caption{The automatic scan sequence lasts one hour.}
  \label{schedula}
 \end{figure}

Both the devices will be set up to take measurements autonomously throughout the night.
Every night, before starting the measurement, a number of laser shots will be fired with the aim of checking and correcting the LIDAR optics alignment.
The best set up will be obtained by tilting the motorized mirror mount inside the laser box around two axes until the position that maximizes the signal over noise ratio at far distances is found.
After that, a sequence of different measurements will start as shown in Fig.~\ref{schedula}.

At the beginning of the sequence the LIDAR fires and acquires a series of laser shots in the horizontal direction. These measurements can be useful to calculate the value of  $\alpha_{aer}$ at ground level, and then to check the horizontal uniformity at ground and to fix the parameter $L_{aer}$ (that is the inverse of $\alpha_{aer}(0)$) used in the Laser Simulation method (see sect. \ref{method}). Acquisitions in this position are also used to interpolate the aerosol optical depth in the zone where the overlap function is not complete and, if the condition of horizontal uniformity is met, to estimate the distance at which the overlap function G(r) becomes constant. Horizontal measurements will thus be used to decrease systematic uncertainties that affect both elastic and Raman analyzes.

Then 200 vertical laser shots are fired vertically at 4 Hz for side-scattering measurements with the AMT.  After each AMT acquisition 120 LED calibration shots will be fired to calibrate the AMT camera. 
The laser is then fired at 100 Hz for twenty minutes for Raman acquisition.  
After two other measurements with the AMT and another twenty minutes Raman acquisition, the LIDAR is tilted at discrete positions in order to perform a multi-angle analysis with the elastic channel. 
A systematical comparison of LIDAR techniques (multi-angle and Raman analysis) allows us to study the uncertainties associated to each method and, in addition, Raman acquisitions at different angles will allow us to quantify the uncertainties associated to the horizontal uniformity hypothesis. LIDAR measurements allow also the evaluation of the PBL height, so these data will be useful to test the improvement obtained introducing a third parameter in the Laser Simulation method.

The sequence ends with 200 shots at 4~Hz inclined at certain angle towards the AMT, so side scattering methods will have to be modified to also analyze data acquired in this condition. A comparison of these results with those obtained with vertical shots  will be used to prove the validity of the horizontal homogeneity assumption. 

ARCADE will also measure the cloud coverage. Clouds can block the UV light, causing a dip in the light profile, or reflect more light in the field of view of the detectors, enhancing the collected light. Measurements of cloud cover, altitude, and cloud optical properties can be performed with both devices. In particular, the LIDAR can identify multiple cloud layers. Information from vertical and inclined shots will be compared with satellite data.
 
Finally, the study of tropospheric aerosol stratification and their variability on short (hourly) and long (seasonal) time scales that will be carry out by ARCADE can be useful in atmospheric physics research field due to the key role that aerosols play in atmospheric radiative processes, phenomenon that have a considerable effect on climate.
Nowadays, there are still no models able to take into account the complex variability of aerosols both in space (on a 10 km scale) and time (on a hourly scale). As a
consequence, the scientific community deeply feels the need to enhance the network of systematic observations in order to reach a clear agreement with satellite
observations and to develop more comprehensive and predictive models.

\section{Conclusions}
ARCADE is a three years project, funded by the italian Ministero dell'Istruzione, dell'Universit\`a e della Ricerca (MIUR), that started in 2012. After the completion of construction of the LIDAR, first measurement campaign will take place first in Turin (urban environment), then in L'Aquila (rural environment) at the end of 2013. Then the LIDAR will be shipped to Lamar (desertic environment) where the device will acquire data together with the existing facility AMT for a year. Results obtained by the project will be included in a database of free access to the scientific community.

\section{Acknowledgments}
We are very grateful to the Ministero dell'Istruzione, dell'Universit\`a e della Ricerca (MIUR), the University of Naples Federico II, and the University of Turin, which are financing this project. We would like to thank the Istituto Nazionale di Fisica Nucleare (INFN), its designers and technicians which are contributing to this project in the design and construction of the Raman lidar. A special thank to the Pierre Auger Collaboration which is actively supporting the ARCADE project. Finally, we would like to thank Dr. Vincenzo Rizi, Dr. Marco Iarlori (CETEMPS - Universit\`a degli Studi dell'Aquila) and Prof. Nicola Spinelli (Universit\`a Federico II di Napoli) for their precious help in the conceptual design of the Raman lidar.

\end{document}